\documentclass[12pt,preprint]{aastex}

\begin{document}

\title{Off-Center Carbon Ignition in Rapidly Rotating, Accreting Carbon-Oxygen White Dwarfs}

\author{Hideyuki Saio}
\affil{Astronomical Institute, Graduate School of Science,
    Tohoku University, Sendai, 980-8578, Japan}
\email{saio@astr.tohoku.ac.jp}

\and

\author{Ken'ichi Nomoto}
\affil{Department of Astronomy, Graduate School of Science, University of Tokyo,
Hongo 7-3-1, Bunkyo-ku, Tokyo 113-0033, Japan}
\email{nomoto@astron.s.u-tokyo.ac.jp}

\begin{abstract}
We study the effect of stellar rotation on the carbon ignition in
a carbon-oxygen white dwarf accreting CO-rich matter.
Including the effect of the centrifugal force of rotation,
we have calculated evolutionary models up to the carbon ignition
for various accretion rates.
The rotation velocity at the stellar surface is set to be
the Keplerian velocity.
The angular velocity in the stellar interior %has been  
is determined by
taking into account the transport of angular momentum 
due to turbulent viscosity.
We have found that an off-center carbon ignition occurs even when
the effect of stellar rotation is included if the accretion rate
is sufficiently high; the critical accretion rate for the off-center
ignition is hardly changed by the effect of rotation.
Rotation, however, delays the ignition;
i.e., the mass coordinate of the ignition layer and the mass of the white dwarf
at the ignition are larger than those for the corresponding no-rotating
model.
The result 
supports
our previous conclusion that a double-white dwarf
merger 
would not
be a progenitor of a SN~Ia.
\end{abstract}


\section{Introduction}

Mass accretion onto a white dwarf, which may occur in a close binary system,
plays an important role in its evolution.
If the accretion rate is sufficiently high, heating of the outer layers 
of the accreting star leads to the ignition of nuclear burning.
A C-O white dwarf accretes CO-rich matter if the companion is also 
a C-O white dwarf (i.e., double C-O white dwarf system).
Whether the carbon is ignited in the stellar center or in the envelope
of an accreting C-O white dwarf
is crucial for the star's fate.
If the ignition occurs at the center, it leads to a SN~Ia explosion.
If carbon burning is
ignited in the envelope, on the other hand, 
the flame propagates inward to the center
converting the C-O white dwarf to an O-Ne-Mg white dwarf \citep{sn85,sn98,twt94} 
that is expected to eventually collapse to a neutron star
\citep{nm84,nm87,nk91}.
\citet{ni85} and \citet{ksn87} have shown that carbon-ignition occurs 
in an outer layer if the C-O white dwarf accretes matter more rapidly than 
$2.7\times10^{-6}M_\odot{\rm y}^{-1}$.

When a double C-O white-dwarf system coalesces due to angular momentum loss
by gravitational wave radiation and/or by a magnetic field, 
the less massive white dwarf is disrupted to
become a thick disk around the more massive white dwarf \citep{it84,web84,
benz90,rs95}.
If the thick disk is turbulent, which is likely, the timescale of viscous angular
momentum transport is very short \citep{ml90} so that
the more massive white dwarf is expected to accrete CO-rich 
matter at a rate close to the Eddington limit from the thick disk.
Therefore, when a double C-O white-dwarf system coalesces, 
carbon burning is ignited in the envelope of the more  massive white dwarf 
to transform it to an O-Ne-Mg white dwarf, thus preventing
the occurrence of the SN~Ia explosion postulated by \citet{it84} and 
\citet{web84}.
In the previous investigations on the carbon ignition, however, 
the effect of rotation was
neglected despite the fact that the accreted matter should bring angular 
momentum to the white dwarf.
It is important to see whether the above conclusion on the fate of a 
double white dwarf merger is unchanged when
the effect of rotation is included.

Recently, \citet{pgit03} investigated the effect of rotation 
on the evolution of white dwarfs that accrete CO-rich matter.
They assumed that all the angular momentum associated with the accreted matter
was brought into the white dwarf. 
They claim that the combined effects of accretion
and rotation induce expansion to
make the surface zone gravitationally unbound and hence suppresses
further accretion in the double white dwarf merger.
They argued that the above effect makes the accretion rate smaller
than the critical value for the occurrence of the off-center carbon ignition, 
and hence the white dwarf can grow up to the Chandrasekhar mass to become a SN~Ia.
In addition to the above assumption
\citet{pgit03b} 
took into account the angular momentum loss via gravitational wave radiation, and
obtained an evolutionary model leading to 
a central carbon ignition. % with reduced accretion rates.

In contrast
to their assumption, 
\citet{pac91} and \citet{pn91} had shown that
when the surface rotation rate is nearly critical,
angular momentum is transported from the star back to the  
accretion disk (see also \citet{fuj95}) and
the star can continue to accrete matter as long as matter exists around the star.
This indicates that we should let matter accrete onto the white dwarf keeping
the surface rotation velocity nearly critical. 
Those results were obtained by considering a central star 
at contact with a thin disk, in contrast of a thick disk expected 
in the case of a merging two white dwarfs.
Although the physics of thick disk accretion is poorly known,
we expect that the thick disk is turbulent and the angular momentum
would be exchanged as fast as in a thin disk.
Therefore, it is likely that the results of \citet{pac91} and \citet{pn91}
are applicable also to thick disk accretion.
Based on this consideration, we assume that
even after the accreting star is spun up
close to the Keplerian velocity at the surface,   
the star accretes matter at the initial rate while the
surface velocity remains equal to the Keplerian velocity.
We note that under this assumption 
the total angular momentum of the accreting white dwarf can even decrease.

Recently, \citet{yl04} computed 
white-dwarf models accreting CO-rich matter, taking into account
the effect of rotation. 
The accretion rates they considered are lower than 
those adopted in our calculations. 
Their assumption at the surface of the accreting white dwarf 
is somewhat different from ours.
They assumed that the matter accretes
onto the white dwarf without bringing angular momentum
when the rotation velocity at the surface of the white
dwarf exceeds the Keplerian velocity.
Under their assumption 
the total angular momentum of the white dwarf never decreases and
is higher than that of our models for a given mass.

\section{Models} \label{models}

The initial models, carbon-oxygen white dwarfs of $0.8M_\odot$ and $1M_\odot$,  
were obtained from zero-age helium main-sequence models 
by artificially suppressing nuclear reactions and converting
chemical compositions to $(X_{\rm C},X_{\rm O},X_{\rm h}) = (0.48,0.50,0.02)$,
where $X_{\rm h}$ denotes the mass fraction of the elements heavier 
than oxygen, and
by letting cool until the central temperature became $\sim10^7$K.  
Table \ref{tab_initialmodel} lists physical characteristics 
of the initial models.
Adopted input physics are mostly the same as those used in Saio \&
Nomoto(1998):
OPAL opacity tables \citep{opal}, neutrino emission rates by
\citet{itoetal89}, and carbon-burning reaction rates by
\citet{cf88}.

\placetable{tab_initialmodel}

Keeping our evolutionary models spherical symmetric,
we have included the effect of rotation only in 
the equation of hydrostatic equilibrium as
\begin{equation}
{1\over\rho}{dP\over dr}= -{GM_r\over r^2} +{2\over3}r\Omega^2,
\end{equation}
where $\Omega$ is the angular velocity of rotation.
The factor $2/3$ comes from taking the mean of the radial component
of the centrifugal force on a sphere \citep{kmt70}.
Spatial and time variation of $\Omega$ is determined
by solving the conservation equation of angular momentum 
\begin{equation}
\left[\partial(r^2\Omega)\over\partial t\right]_{M_r}
={\partial\over\partial M_r}
\left[(4\pi\rho r^3)^2\nu{\partial\Omega\over\partial M_r}\right],
\end{equation}
where $\nu$ stands for turbulent viscosity. 
The outer boundary condition $\Omega=\Omega_{\rm K}$ was
imposed at the photosphere, where 
$\Omega_{\rm K} = \sqrt{GM/R^3}$ is the Keplerian 
angular velocity at the stellar radius $R$. 
We have neglected the rotation of the initial models.

We have adopted two kinds of formulas for the turbulent viscosity;
one from \citet{fuj93} and another from \citet{zahn92}.
The turbulent viscosity formulated by \citet{fuj93} includes
the effect of baroclinic instability as well as shear instability.
We will refer this case as case A; the viscosity is given by
\begin{equation}
\nu_{\rm A} = \nu_{\rm KH} + \nu_{\rm BC},
\end{equation}
where
\begin{equation}
\nu_{\rm KH} = \left\{
\begin{array}{lll}\displaystyle
{1\over2 Ri^{1/2}}(1-4Ri)^{1/2}H_p^2N & {\rm for} & Ri < 1/4 \cr
0 & {\rm for} & Ri \ge 1/4; \cr
\end{array}
\right.
\end{equation}
\begin{equation}
\nu_{\rm BC} = \left\{
\begin{array}{lll}\displaystyle
{1\over3 Ri^{1/2}}H_p^2\Omega & {\rm for} & Ri \le Ri_{\rm BC} \cr
\displaystyle
{1\over3 Ri^{3/2}}H_p^2\Omega & {\rm for} & Ri > Ri_{\rm BC},
\end{array}
\right.
\end{equation}
\begin{equation}
Ri = N^2\left(d\Omega\over d\ln r\right)^{-2}, \qquad
Ri_{\rm BC} = 4\left({r\over H_p}\right)^2{\Omega^2\over N^2}
\end{equation}
with $N$ being Brunt-V\"ais\"al\"a frequency defined as
$N^2=g(\Gamma_1^{-1}d\ln P/dr - d\ln\rho/dr)$ and $H_p$ 
the pressure scale height.

For the second case, which is referred to as case B, we have included turbulent 
viscosity due to vertical shear instability formulated by \citet{zahn92}
but neglected the effect of meridional circulation; 
we have used the following equation,
\begin{equation}
\nu_{\rm B} = {2\over45} {K\over N^2}\left(d\Omega\over d\ln r\right)^2,
\end{equation}
where $K$ is the thermal diffusivity defined by 
$K = 4acT^3/(3\kappa\rho^2C_p)$.
As we will see in the next section, the viscosity in case A is so efficient
in transporting the angular momentum that the rotation becomes nearly 
uniform in the white-dwarf interior.
In case B, on the other hand, the angular momentum
is slowly transported into the deep interior. 
We assume that the two models for the transport of angular momentum 
bracket what occurs in a real star.

\section{Numerical Results} \label{results}

\begin{figure}
\plotone{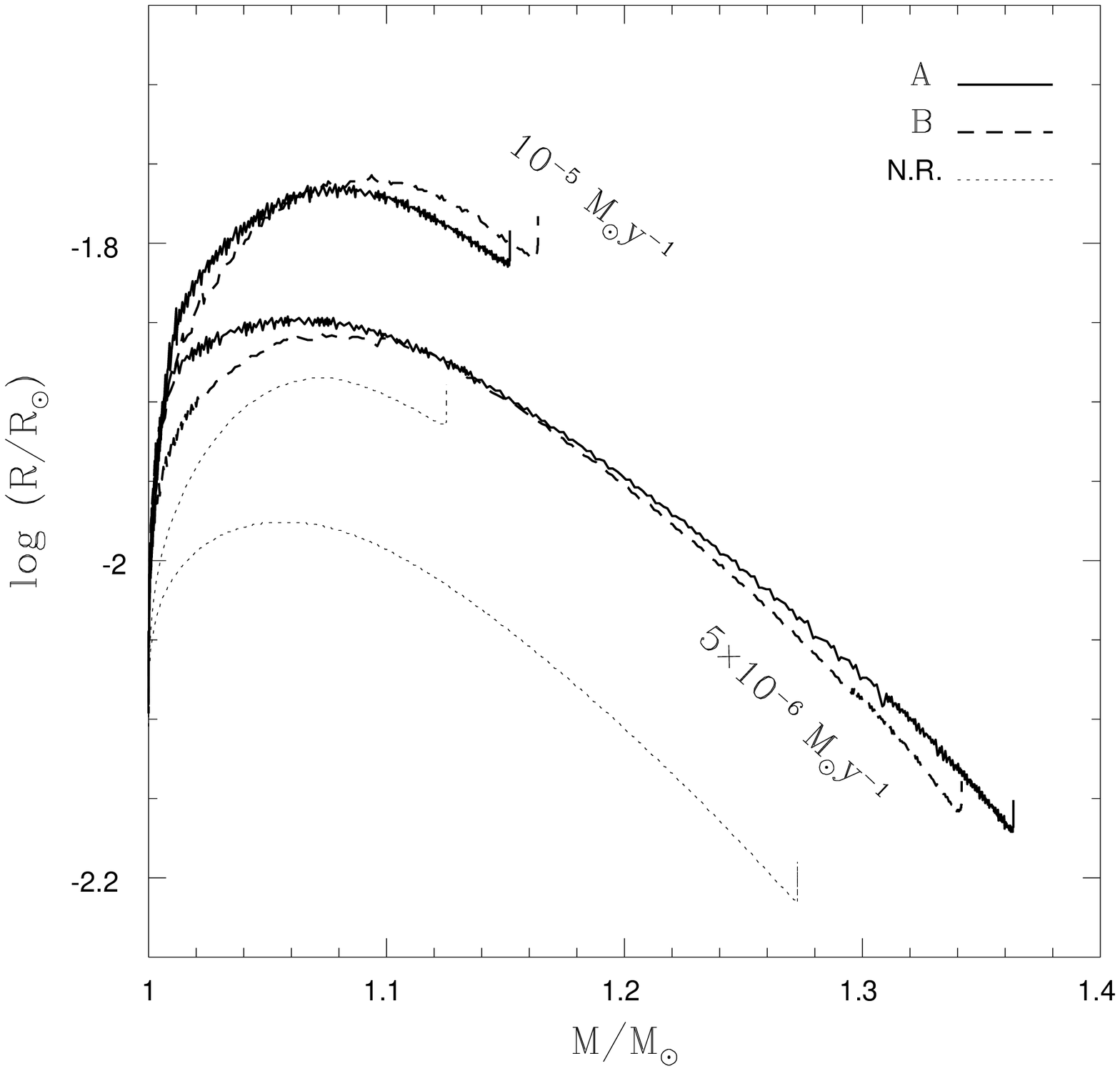}
\caption{Evolution of C-O white dwarfs that accrete CO-rich matter at rates of 
$10^{-5}$ and $5\times10^{-6}M_\odot$y$^{-1}$.
The ordinate and the abscissa are, respectively, the radius and the mass
of the white dwarf.
The solid (case A) and dashed (case B) lines are for rotating models in which the 
angular momentum transport is calculated using relatively 
high (A) and low (B) turbulent viscosities, respectively.
The dotted lines are for non-rotating models.
The nearly vertical increase in radius at the end of each evolution
indicates the occurrence of an off-center carbon ignition. 
}
 \label{mass_radius}
\end{figure}

\placefigure{mass_radius}

Figure \ref{mass_radius} shows the evolution of radius as a function 
of the total mass for C-O white-dwarfs that accrete CO-rich matter at
rates of $10^{-5}$ and $5\times10^{-6}M_\odot$y$^{-1}$
from the beginning of accretion to the  off-center carbon ignition.
The initial mass of the models is $1M_\odot$.
Wiggles seen in this figure are numerical artifacts.
The dotted lines show the evolution of non-rotating models,
the solid lines for rotating models in case A, 
and the dashed lines for rotating models in case B.

In the early phase of accretion the radius of the white dwarf
increases rapidly. This is caused by the fact that 
its envelope is heated by the accretion,
and the centrifugal force of rotation lifts the envelope.
The initial rapid expansion in case A models slows down
when the angular momentum begins to be transferred back to the
accretion disk (Fig.\ref{ev_angmom} below). 
Such transition to a slower evolution did not occur in the models
of \citet{pgit03b}, because they did not consider the possibility
of transferring the angular momentum back to the disk, but did
reduce the accretion rate.
In later phase of accretion, 
the radius begins to decrease.
The decrease in radius is caused by the increase of the white-dwarf mass.
During this phase the total angular momentum is nearly constant
or increases gradually. %(see Fig. \ref{ev_angmom} below).
The lifting effect of rotation makes the radius of a rotating model larger
than that of the non-rotating model with the same mass and the accretion rate.

A short vertical increase in radius at the end of each evolution track
indicates the occurrence of the off-center carbon ignition.
For a given accretion rate, the off-center carbon ignition 
occurs later in a rotating model than in the non-rotating model;
i.e., the mass of the rotating white dwarf at the carbon ignition
is larger than that for the corresponding non-rotating model.

\begin{figure}
\plotone{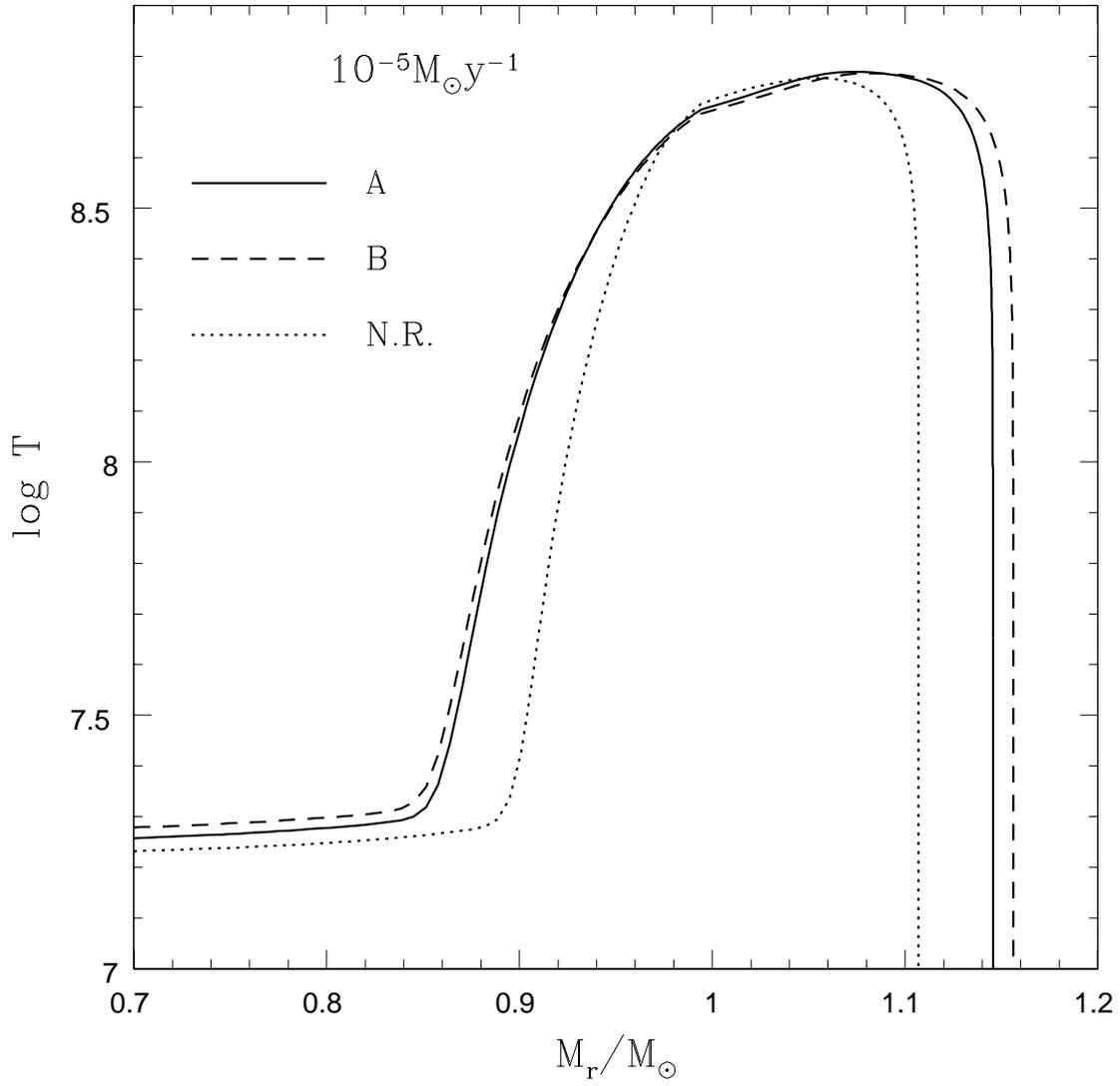}
\caption{Temperature distributions in models close to the off-center
carbon ignition for an accreting rate of $10^{-5}M_\odot$y$^{-1}$. 
The solid and the dashed line are for rotating models with different
turbulent viscosities. The dotted line is for the 
non-rotating model.
}
\label{mr_t}
\end{figure}

\placefigure{mr_t}

Figure \ref{mr_t} shows the temperature versus $M_r$ in the interior of
models close to the off-center carbon ignition.
These models have evolved from the initial model of $1M_\odot$ with
an accretion rate of $10^{-5}M_\odot$y$^{-1}$.
The solid and dashed lines are for rotating models, and
the dotted line for the non-rotating model.
The outer layers consisting of accreted matter ($M_r > 1M_\odot$) have high
temperatures heated by the gravitational energy release.
The heat is conducted into the layers of the original white dwarf.
The heat-conduction front, however, has not advanced much 
during the time between the start of accretion and the off-center 
ignition ($\sim 10^4$y). 
Rotating models have broader temperature peaks than the non-rotating model,
because they accrete more mass and hence the time from the start of accretion
to the carbon ignition is longer.
The core temperature has increased from the initial value $\log T =7.08$
due to compression.

Table \ref{tab_sum} lists the total mass $M_{\rm ig}$ and the
central density $\rho_{\rm c}$ of the model in which 
off-center carbon-ignition occurs, and the mass interior to
the ignition layer $M_r({\rm ig})$ for each case computed, 
where the values of masses are given in units of $M_\odot$, 
$\dot M$ means accretion rate in units of $M_\odot$y$^{-1}$,
and $M_{\rm i}$ stands for the initial mass. 
The difference in $M_{\rm ig}$ between a rotating and the non-rotating
model for a given accretion rate can be as large as $\sim 0.1M_\odot$.
The value of $M_r({\rm ig})$ is also larger in the rotating model.
Rotation delays the carbon ignition because
the centrifugal force reduces the effective gravity to lift
the outer layers and reduce the temperature there \citep{pgit03}.
The lifting effect of rotation  appears also in the central density,
$\rho_c$, which is lower in the rotating model than in the non-rotating
model with a similar mass.
The case A models with $\dot M = 3\times 10^{-6}M_\odot$y$^{-1}$
did not ignite carbon. 
They began to expand before the maximum temperature
in the outer layer reaches to the ignition temperature.

Table \ref{tab_sum} shows that the results hardly depend on the 
initial mass of the white dwarf. 
The results confirm essentially the conclusion of the previous work 
done without including the effect of rotation that carbon is ignited
in the outer layer of a C-O white dwarf if the accretion rate is
fast enough.

\placetable{tab_sum}

\placefigure{rel_omega}

\section{Discussion}\label{discussion}

\begin{figure}
\plotone{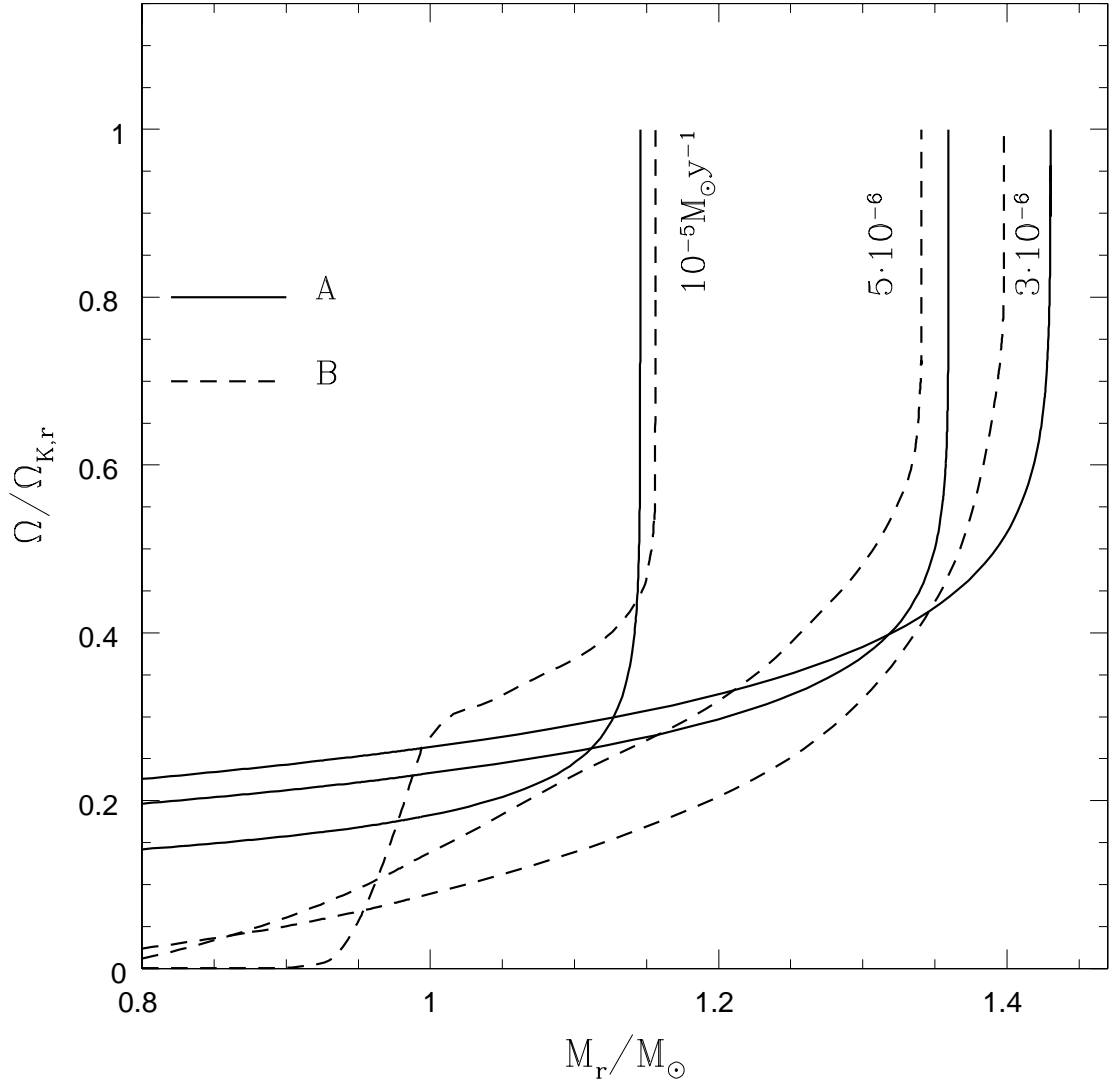}
\caption{Runs of the ratio of angular velocity of rotation to the
local Keplerian frequency, $\Omega_{{\rm K},r}$, 
in the interior of models for accretion
rates of $10^{-5}$, $5\times10^{-6}$, and $3\times10^{-6}M_\odot$y$^{-1}$.
The abscissa represents the mass coordinate in units of solar mass.
}
 \label{rel_omega}
\end{figure}

\begin{figure}
\plotone{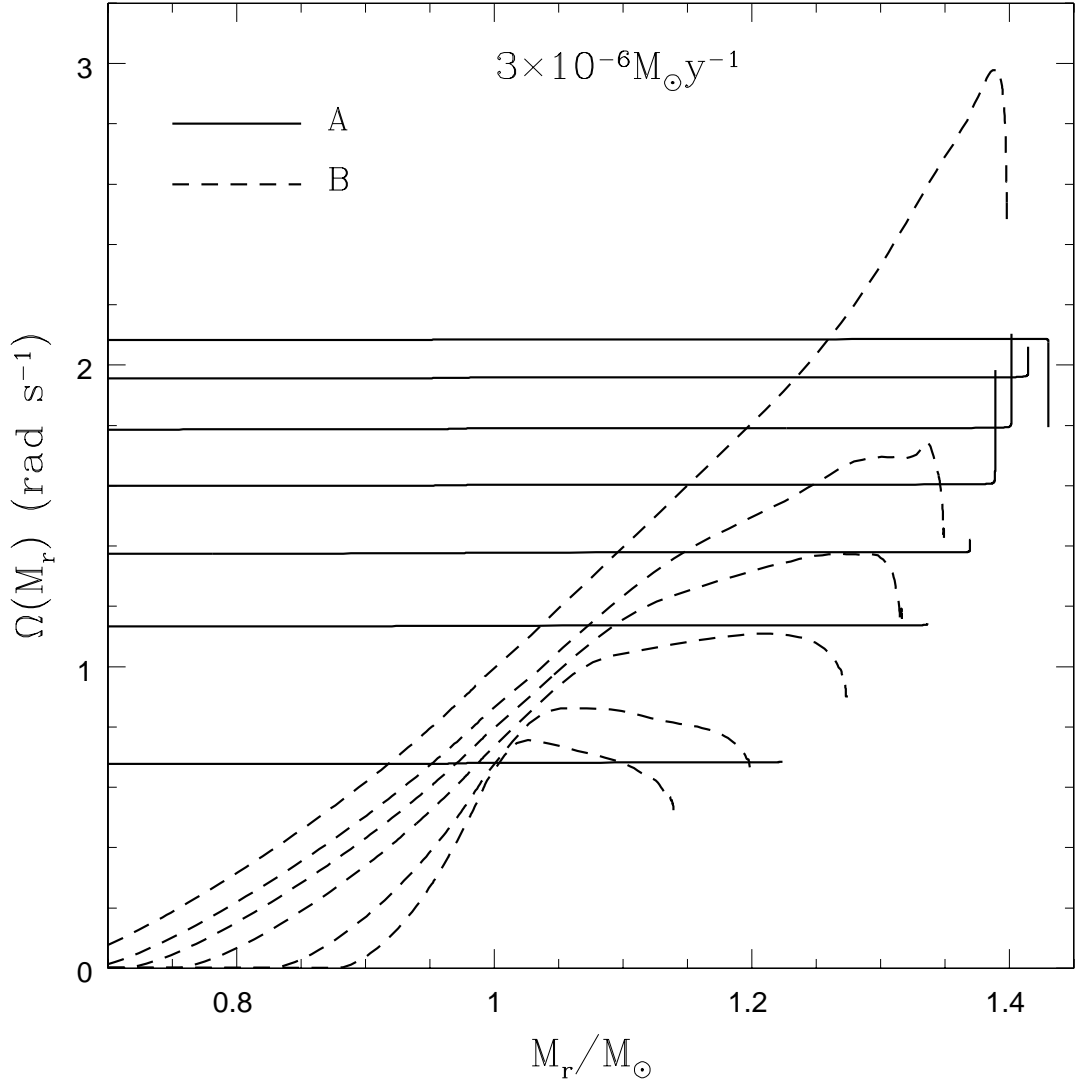}

\caption{
Runs of angular velocity of rotation, $\Omega(M_r)$,
in the interior of models at selected evolutionary stages for the  
accretion rate $\dot M = 3\times10^{-6}M_\odot{\rm y}^{-1}$.
}
\label{ang_freq_evol_3em6}
\end{figure}

\begin{figure}
\plotone{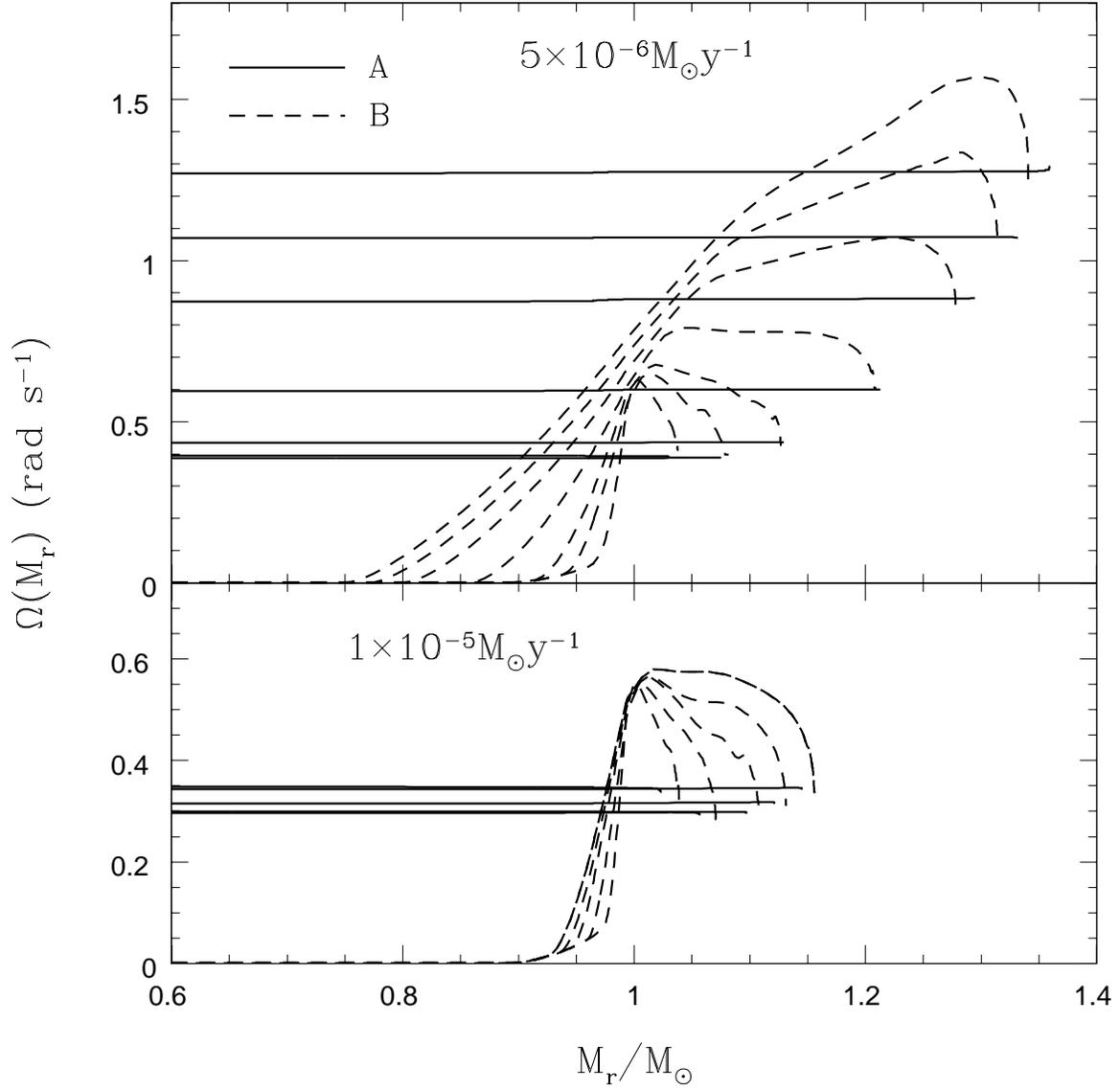}

\caption{
The same as Fig. \ref{ang_freq_evol_3em6} but for
for $\dot M = 10^{-5}$ (lower panel) 
and $5\times10^{-6}M_\odot{\rm y}^{-1}$ (upper panel).
}
\label{ang_freq_evol}
\end{figure}

Figure \ref{rel_omega} shows runs of the ratio of angular velocity
to its local critical value, $\Omega/\Omega_{{\rm K},r}$, in the
models at evolutionary stages around the carbon ignition 
for $\dot M = 10^{-5}$, $5\times10^{-6}$ and $3\times10^{-6}M_\odot {\rm y}^{-1}$.
(For the case A of $\dot M = 3\times10^{-6}M_\odot {\rm y}^{-1}$,
$\Omega/\Omega_{{\rm K},r}$ in the last model computed is shown.)
Here $\Omega_{{\rm K},r} = \sqrt{GM_r/r^3}$ is the local Keplerian velocity.
%Shown are the case A and case B models 
%with accretion rates of 
%$\dot M = 10^{-5}$, $5\times10^{-6}$ and $3\times10^{-6}M_\odot {\rm y}^{-1}$. 
In all cases, $\Omega/\Omega_{{\rm K},r}$ increases rapidly toward the surface
in the very superficial layer, 
which is due to a rapid increase in the radius of mass shell
toward the surface.
In case A the viscous transport of angular momentum is very efficient
so that the rotation becomes nearly uniform
(Figs. \ref{ang_freq_evol_3em6},\ref{ang_freq_evol}). 
Therefore, the %dashed 
solid curves in Figure \ref{rel_omega} are almost parallel to  
the runs of $r^{3/2}M_r^{-1/2}$.
An exception is the outermost layer of the case A model for 
$\dot M = 3\times10^{-6}M_\odot {\rm y}^{-1}$. 
Since the radius of the model is rapidly increasing and hence
the surface angular velocity (= Keplerian velocity) is 
rapidly decreasing, there is a steep outward decrease in the angular velocity
of rotation in the outermost layer (Fig.\ref{ang_freq_evol_3em6} below).

In case B
the angular momentum is hardly transported into
the original white dwarf ($M_r \le 1M_\odot$).
Thus, the angular velocity at $M_r \sim 1M_\odot$
in the case B model of $\dot M=10^{-5}M_\odot{\rm y}^{-1}$
is considerably larger than in the corresponding case A model.
This account for the result that
$M_{\rm ig}$ and $M_r({\rm ig})$ in case B with 
$\dot M = 10^{-5}M_\odot{\rm y}^{-1}$ are slightly larger than those for case A.
In other cases, however, the effect of rotation is generally stronger
in case A models than in case B models.

\placefigure{ang_freq_evol_3em6}
\placefigure{ang_freq_evol}

Figure \ref{ang_freq_evol_3em6} shows runs of
angular velocity of rotation $\Omega(M_r)$ in the interior of models 
at selected phases of evolution for $\dot M = 3\times 10^{-6}M_\odot$y$^{-1}$
($M_{\rm i}= 1M_\odot$).
Figure \ref{ang_freq_evol} shows the same information
for $\dot M = 1\times 10^{-5}$ (lower panel) and 
$5\times10^{-6}M_\odot{\rm y}^{-1}$ (upper panel).
Obviously, the angular momentum transport by the 
turbulent viscosity in case A is so efficient that 
$\Omega(M_r)$ is nearly constant in the interior. 
In the case A  model with $\dot M = 3\times 10^{-6}M_\odot$y$^{-1}$, 
$\Omega(M_r)$ varies steeply in the very superficial layer 
when the total mass is larger than $1.35M_\odot$
(Fig. \ref{ang_freq_evol_3em6}). 
Since $\Omega$ at the stellar surface is always set to be the critical 
frequency $\Omega_K$ ($\propto M^{1/2}R^{3/2}$),
a steep increase in $\Omega(M_r)$ toward the surface indicates
the star is rapidly contracting, and a steep decrease indicates a 
rapid expansion. 
The surface angular velocity changes so rapidly that
the transport of angular momentum is not efficient enough
to keep the angular velocity of rotation in the interior close to
the surface value.
The model  contracts first and then expands rapidly.

In the interior of a case B model there is a peak in $\Omega(M_r)$, which
is produced by compression.
The angular velocity at the peak is higher than that at the surface of the model.
Although $\Omega$ decreases steeply outward near the surface,
the specific angular momentum $r^2\Omega$ increases outward there because 
the distance from the center of a mass shell increases outward 
steeply near the surface.
In early phases of accretion, there is a narrow layer where 
the specific angular momentum decreases outward
just outside of the peak in $\Omega$ at $M_r\sim 1.0M_\odot$.
But, the structure is always stable 
to the Solberg-H\o iland instability, because the temperature gradient
is positive there and hence the stabilizing effect of buoyancy exceeds
the destabilizing effect of the negative gradient of the specific angular 
momentum distribution.

\begin{figure}
\plotone{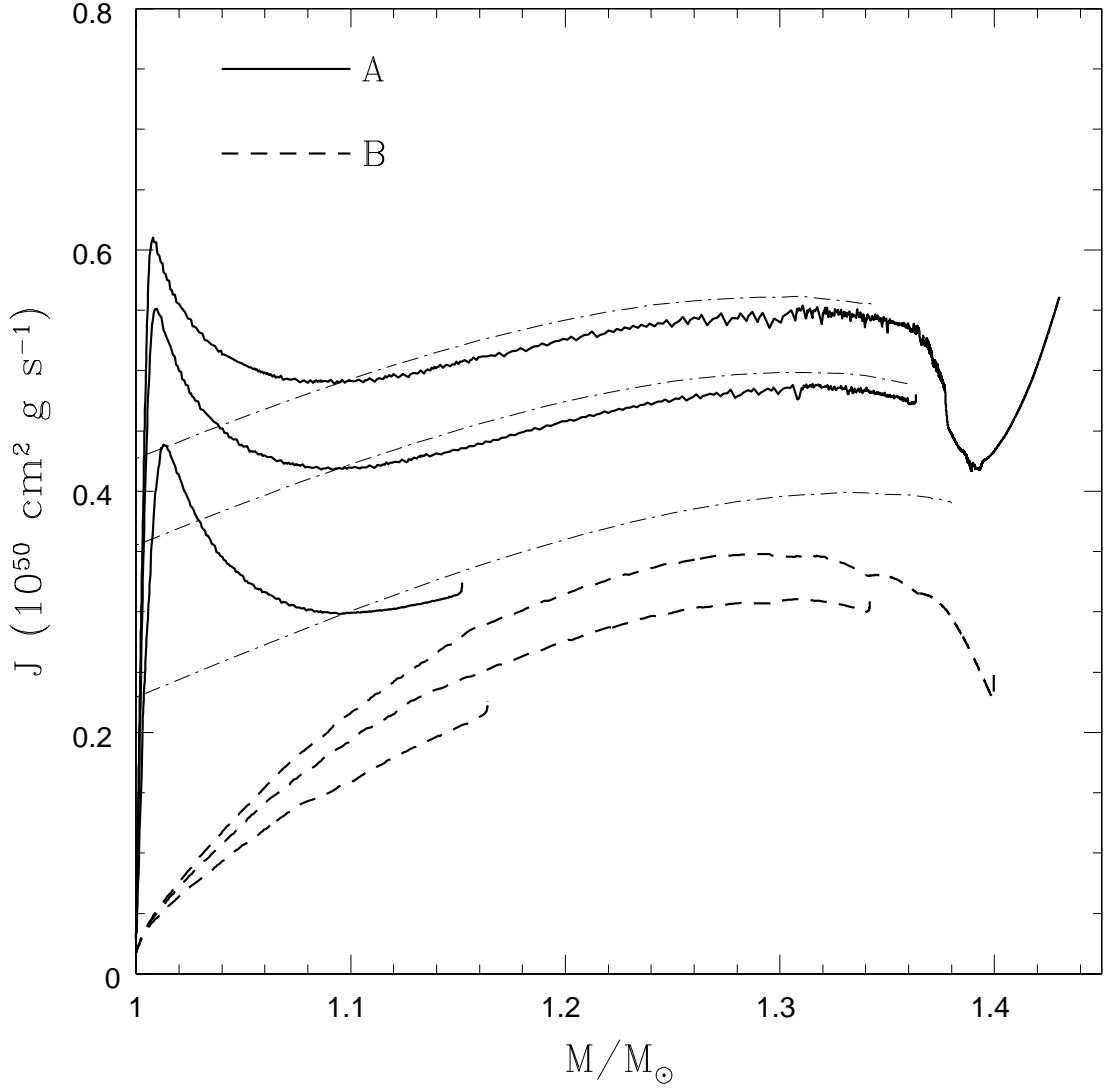}
\caption{
Evolution of the total angular momentum $J$ of C-O white dwarfs
with respect to the total mass
for accretion rates 
of $3\times10^{-6}$, $5\times10^{-6}$, and $10^{-5}M_\odot$y$^{-1}$. 
The lower the accretion rate, the higher is the total angular momentum
for a given total mass. 
Dash-dotted lines show the relations of uniformly and critically 
rotating steady models 
for those mass-accretion rates.
}
\label{ev_angmom}
\end{figure}

\placefigure{ev_angmom}

Figure \ref{ev_angmom} shows the evolution of the total angular
momentum of the models accreting matter at
$\dot M = 10^{-5}$, $5\times10^{-6}$ and $3\times10^{-6}M_\odot {\rm y}^{-1}$.
The higher the accretion rate, the lower the total angular momentum 
at a given total mass.
In case A models %with larger viscosity 
(solid lines) the total 
angular momentum, $J$, increases rapidly until a mass of $\approx 0.01M_\odot$ 
is accreted.
After the total angular momentum attains a peak, (when the angular velocity 
in the interior becomes almost constant), it decreases as the total mass
increases to $\approx 1.1M_\odot$, returning part of the accumulated 
angular momentum to the disk.
For $M \ga 1.1M_\odot$, $J$ increases but very slowly.
In this phase most of the angular momentum associated with the accreted matter
is returned to the disk.

To compare the evolutionary $M - J$ relations with those of simpler models, 
we have computed uniformly and critically rotating steady models, 
in which the rate of entropy change $(\partial s/\partial t)_{M_r}$
is replaced with $-(\partial s/\partial \ln q)_t\dot M/M$ where $q=M_r/M$
(e.g., \citet{ksn87}),
and the nuclear reactions are artificially suppressed.
The $M-J$ relation of the steady model for each accretion rate is 
shown by a dash-dotted line in Fig \ref{ev_angmom}. 
(Detailed properties of the steady models will be presented in a 
future paper.)
This figure shows that the gradual increase of $J$ with a increase in mass
in the range $M \ga 1.1M_\odot$ occurs along the 
$M - J$ relation of the steady model with the same accretion rate,
indicating the evolution in this phase occurs nearly homologously.

The total angular momentum in a case B model 
is smaller than the case A model with the same mass and accretion rate.
Since the angular momentum transport in the case B models is less efficient
then in the case A models, more angular momentum is returned to the
accretion disk in the former case.

The total angular momentum of a case A model is smaller than 
that of the uniformly and critically rotating white dwarf of the same mass 
without accretion (Uenishi, Nomoto, Hachisu 2003).
The difference is attributed to the fact that the critical angular
velocity of rotation for the 
the accreting white dwarf is lower than that of the white dwarf without
accretion, because the radius of the former is larger.

When the mass of the white dwarf becomes larger than $\sim 1.36M_\odot$, 
the total angular momentum 
$J$ of the case A model of $\dot M = 3\times10^{-6}M_\odot$y$^{-1}$
begins to decrease due to a rapid decrease in radius.
The commencement of the rapid contraction roughly corresponds
to the phase when the mass of the white dwarf exceeds
the maximum mass of the steady model for the same accretion rate.
At $M\approx1.39M_\odot$ the radius and the total angular momentum
become minimum and carbon burning is almost ignited at $M_r = 1.38M_\odot$.
However, at this point the model begins to expand and the total
angular momentum begins to increase. 
Due to the expansion, the maximum temperature in the interior 
decreases, and  carbon ignition does not occur in this model.
Further computation became too difficult to continue.
We note however that if the viscosity is switched to the case B viscosity,
we can continue the calculation to have a massive (say $2M_\odot$) model 
having a high angular momentum, in which $\Omega(M_r)$ has a broad peak 
similar to those shown in \citet{yl04}.

Finally, we note that the ratio of the rotational kinetic energy
to the absolute value of the gravitational potential energy defined as
\begin{equation}\label{tw}
{T\over |W|} = {1\over2}\int_0^Mr^2\Omega^2dM_r\left[\int_0^M {GM_r\over r} dM_r
\right]^{-1}
\end{equation}
is always less than $0.03$ in our models.
For such a small ratio, no dynamical instability occurs unless
a very strong differential rotation in the deep interior
of the star exists \citep{ske03}. 
The ratios $T/|W|$ of our models are smaller than the corresponding
ratio denoted by $\gamma$ in \citet{pgit03b} for the critically
rotating model with a similar accretion rate, where $\gamma$ is defined as
\begin{equation}\label{gamma}
\gamma = {1\over2}\int_0^Mr^2\Omega^2dM_r\left[{1\over2}{GM^2\over R}\right]^{-1}.
\end{equation}
A rotating and accreting white dwarf has a large radius because
the outer envelope (in which only a small fraction of mass resides) 
is expanded, 
while the inner layers contribute to the gravitational energy. 
Therefore, $|W|$ is generally larger than ${1\over2}GM^2/R$ for the rotating
and accreting models.
In fact,
$|W|$ of our models can be as large as
$1.6\times GM^2/R$, which makes $T/|W|$ as small as $\gamma/3$. 
Even if $\gamma$ defined by equation (\ref{gamma})
reached $0.14$, $T/|W|$ can be as small as $\sim 0.05$.
Since $T/|W|$ should be used for the stability criterion rather than
$\gamma$, our models should be stable.
We also note that the $T/|W|$ ratios of the models in \citet{yl04}
are larger than those of our models.
This is attributed to the fact that in the interior of
their models there exist a broad peak of the angular velocity.

\section{Conclusion}

Taking into account the centrifugal force of rotation, 
we computed models of CO-white dwarfs that accrete CO-rich matter
up to the carbon-ignition.
We assumed that the surface angular velocity of rotation is equal to 
the Keplerian angular velocity. 
We have found that carbon is 
ignited in an outer layer if the accretion rate is
greater than 
$3\times10^{-6}M_\odot{\rm y}^{-1}$. 
The rotation delays the ignition. 
The difference in the total mass at the ignition between rotating
and non-rotating models can be as large as $\sim 0.1M_\odot$,
depending on the assumed turbulent viscosity
and the accretion rate (see \citet{unh03} for
the 2D models).

When the double white dwarfs coalesce, the more massive white
dwarf is expected to accrete CO-rich matter at a rate
close to the Eddington limit ($\sim 10^{-5}M_\odot$yr$^{-1}$).
We have found in this paper that such a high accretion rate 
leads to an off-center carbon ignition even if
the effect of rapid rotation is included.
Once the ignition occurs, the carbon flame propagates through the interior
to the stellar center due to conduction and the C-O white dwarf
is converted to an O-Ne-Mg white dwarf peacefully 
\citep{sn85,sn98,twt94}.
The result of our computation including the effect of rotation
supports
our previous conclusion 
that a double C-O
white dwarf merger would {\it not} yield a SN Ia.

\acknowledgments

We are grateful to the referee for the constructive comments.
This work has been supported in part by the grant-in-Aid for
COE Scientific Research (15204010, 16042201, 16540229) of the Ministry of
Education, Culture, Science, Sports, and Technology in Japan.

\clearpage
\clearpage

\clearpage

\clearpage

\clearpage

\clearpage

\clearpage

\begin{deluxetable}{ccccc}
%\tabletypesize{\scriptsize}
\tablecaption{Physical characteristics of initial models}
\tablewidth{0pt}
\tablehead{
\colhead{$M/M_\odot$} & \colhead{$\log T_{\rm c}$}   & 
\colhead{$\log \rho_{\rm c}$}  & \colhead{$\log L/L_\odot$} &
\colhead{$\log R/R_\odot$} 
}
\startdata
 
$0.8$ & $7.086$ & $7.024$ & $-2.286$ & $-1.999$ \cr
$1.0$ & $7.081$ & $7.528$ & $-2.216$ & $-2.105$

\enddata 
\label{tab_initialmodel}
\end{deluxetable}

%\clearpage

\begin{deluxetable}{ccccccccccc}
%\tabletypesize{\scriptsize}
\tablecaption{Summary of numerical results}
\tablewidth{0pt}
\tablehead{
  & & \multicolumn{3}{c}{No rot.} %& \multicolumn{3}{c}{Rigid rot.}
& \multicolumn{3}{c}{Rot. A} & \multicolumn{3}{c}{Rot. B}\cr
\colhead{$\dot M$} & \colhead{$M_{\rm i}$}   & 
\colhead{$M_r({\rm ig})$}  & \colhead{$M_{\rm ig}$} & 
\colhead{$\log\rho_{\rm c}$}  &
\colhead{$M_r({\rm ig})$}  & \colhead{$M_{\rm ig}$} & \colhead{$\log\rho_{\rm c}$} &
\colhead{$M_r({\rm ig})$}  & \colhead{$M_{\rm ig}$} & \colhead{$\log\rho_{\rm c}$} 
}
\startdata

\\
   $3\times10^{-6}$ & 1.0 & 1.35 & 1.36 & 9.15 &  
                            -- & -- & -- & 1.38 & 1.40  & 9.25 \\
   $3\times10^{-6}$ & 0.8 & 1.35 & 1.36 & 9.13 &
                            -- & -- & -- & 1.39 & 1.40  & 9.24 \\

\\
   $5\times10^{-6}$ & 1.0 & 1.24 & 1.27 & 8.47 &  
                            1.33 & 1.36 & 8.64 & 1.31 & 1.34 & 8.67 \\
   $5\times10^{-6}$ & 0.8 & 1.25 & 1.28 & 8.48 & 
                            1.33 & 1.37 & 8.64 & 1.31 & 1.35 & 8.66 \\
\\
   $1\times10^{-5}$ & 1.0 & 1.05 & 1.12 & 7.82 & 
                            1.07 & 1.15 & 7.83 & 1.08 & 1.16 & 7.87 \\
   $1\times10^{-5}$ & 0.8 & 0.98 & 1.09 & 7.65 & 
                            1.01 & 1.13 & 7.70 & 1.02 & 1.14 & 7.72 \\

\enddata

\label{tab_sum}
\end{deluxetable}

\end{document}